\documentclass{article}

\usepackage{arxiv}

\usepackage[utf8]{inputenc} 
\usepackage[T1]{fontenc}    
\usepackage{hyperref}       
\usepackage{url}            
\usepackage{booktabs}       
\usepackage{amsfonts}       
\usepackage{nicefrac}       
\usepackage{microtype}      
\usepackage{lipsum}
\usepackage{graphicx}
\usepackage{amsmath}
\usepackage{bm}
\usepackage{float}
\usepackage[round,authoryear]{natbib}
\graphicspath{ {./images/} }

\title{A two-stage approach to heat-mortality risk assessment comparing multiple exposure-to-temperature models: the case study in Lazio, Italy}

\author{
Emiliano Ceccarelli \\
Department of Statistical Sciences \\
Sapienza University of Rome \\
Piazzale Aldo Moro, 5 \\
00185, Rome, Italy
\And
Jorge Castillo-Mateo \\
Department of Statistical Methods and IUMA \\
University of Zaragoza \\
Pedro Cerbuna 12 \\
50009, Zaragoza, Spain
\And
Sandra Gud\v{z}i\={u}nait\.{e} \\
MRC Centre for Environment and Health \\
Department of Epidemiology and Biostatistics \\
Imperial College London \\
White City Campus \\
W12 0BZ, London, UK
\And
Giada Minelli \\
Statistical Services \\
Istituto Superiore di Sanità \\
Viale Regina Elena, 299 \\
00161, Rome, Italy
\And
Giovanna Jona Lasinio \\
Department of Statistical Sciences \\
Sapienza University of Rome \\
Piazzale Aldo Moro, 5 \\
00185, Rome, Italy
\And
Marta Blangiardo \\
MRC Centre for Environment and Health \\
Department of Epidemiology and Biostatistics \\
Imperial College London \\
White City Campus \\
W12 0BZ, London, UK
}

\begin{document}
\maketitle
\begin{abstract}
This study investigates how different spatiotemporal temperature models affect the estimation of heat-related mortality in Lazio, Italy (2008--2022). First, we compare three methods to reconstruct daily maximum temperature at the municipality level: 1. a Bayesian quantile regression model with spatial interpolation, 2. a Bayesian Gaussian regression model, 3. the gridded reanalysis data from ERA5-Land. Both Bayesian models are station-based and exhibit higher and more spatially variable temperatures compared to ERA5-Land. Then, using individual mortality data for cardiovascular and respiratory causes, we estimate temperature-mortality associations through Bayesian conditional Poisson models in a case-crossover design. Exposure is defined as the mean maximum temperature over the previous three days. Additional models include heatwave definitions combining different thresholds and durations. All models exhibit a marked increase in relative risk at high temperatures; however, the temperature of minimum risk varies significantly across methods. Stratified analyses reveal higher relative risk increases in females and the elderly (80+). Heatwave effects depend on the definitions used, but all methods capture an increased mortality risk associated with prolonged heat exposure. Results confirm the importance of temperature model choice in epidemiology and provide insights for early warning systems and climate-health adaptation strategies.
\end{abstract}


\section{Introduction}
The impact of high temperatures on human health is a well-established public health concern, with numerous studies linking both heat and cold exposure to increased mortality, particularly from cardiovascular and respiratory causes \citep{barnett2007temperature, baccini2008heat, gasparrini2015mortality}. In the context of climate change, heat-related health risks are expected to become more severe, as both the frequency and intensity of extreme temperature events are projected to increase in the coming decades \citep{alfano2023lancet}. 
While the adverse effects of extreme heat are now well recognized, discussion follows on the optimal temperatures that correspond to minimum effects for various health outcomes, which vary for different regions of the World and depend on the temperature database considered.

The effect of heat is typically investigated by regressing death counts within a time frame (usually weeks or days) against a respective measure of temperature (e.g. average or maximum) on a continuous scale. An alternative way to define heat exposure is through the concept of heatwaves, which are periods of multiple consecutive days in which temperatures surpass a certain threshold \citep{Robinson_2001, Perkins_Alexander_2013}. It has been hypothesize that exposure to multiple consecutive hot days may constitute a health risk in addition to daily temperature alone \citep{Gasparrini_Armstrong_2011}. Indeed, the physiological and behavioral mechanisms for thermo-regulation are prone to exhaustion and ultimately failure among the most vulnerable populations (e.g. the elderly and those living with chronic conditions) \citep{Bennett_Blangiardo_Fecht_Elliott_Ezzati_2014, Blum_DeIngeniis_Shill_Stone_Sheffield_Nomura_2025}. Hence, a way to account for this is to add a categorical variable indicating heatwave days to test whether it captures some variations in health outcomes other than continuous temperature alone \citep{Robinson_2001}. This study will employ both methods for testing the effects of heat on mortality.  

Most epidemiological studies rely on gridded satellite-based reanalysis data products, also called gridded climate datasets (GCDs), such as Copernicus' ERA5-Land \citep{munoz2021era5, copernicus2023era5}, to assess the dose-response association between heat and health \citep{ballester2023heat, ceccarelli2024understanding}. GCDs are easily accessible and provide a large temporal coverage, but are also prone to measurement error in complex terrains such as coastal areas, affected by mixed land-sea pixels, or mountainous regions, characterized by large elevation differences; this can affect the accuracy of local temperature estimates \citep{donat2014consistency}. Many argue that ground-based weather stations should be preferred since they provide direct and precise observations at specific locations \citep{mendelsohn2007climate, colston2018evaluating}. However, weather stations are often unevenly distributed, with denser coverage in highly populated areas and in more developed regions.
These differences in spatial resolution, representativeness, and data-generating processes can result in different exposure estimates at the local scale \citep{sheridan2020comparison, mcnicholl2021evaluating}, especially for spatially heterogeneous regions. In turn, such discrepancies can lead to differences in estimated health risks, with important public health implications, for instance around early warning systems for high temperatures. 

The spatial scale of the temperature data plays a crucial role, since it should align with the resolution of the health data, commonly available at the municipality level. While ERA5-Land provides a relatively high resolution of 9 km \citep{munoz2021era5}, this is still insufficient to capture fine-scale temperature variability between neighbouring municipalities, especially in geographically complex regions. In contrast, station-based data can offer highly localized information, but their spatial distribution is inconsistent, depending on factors like population density and geography. This often results in clusters of stations in urban centers and a lack of coverage in rural or mountainous zones. 

In literature, temperature estimates derived from ground-based monitoring stations can be effectively interpolated through spatiotemporal statistical models. For instance, \cite{benavides2007geostatistical} uses a kriging approach comparing several geostatistical and regression-based methods in a mountainous region of northern Spain. \cite{craigmile2011space}, \cite{verdin2015coupled}, and \cite{castillo2022} developed spatiotemporal hierarchical Bayesian models. In particular, \cite{craigmile2011space} estimated daily mean temperatures in Central Sweden using wavelet-based trend components, while \cite{verdin2015coupled} and \cite{castillo2022} modelled minimum and/or maximum temperatures in the Pampas region of Argentina and the Aragón region of Spain, respectively, incorporating autoregressive structures and spatial Gaussian processes.  In a similar context, quantile regression \citep{koenker1978regression, koenker2005quantile}, and quantile autoregression \citep{koenker2006quantile}, have been used extensively to model extreme temperatures, both in a Frequentist \citep{gao2017quantile} and Bayesian approach \citep{ferraz2020bayesian, castillo2023spatial}. 
Yet, a comprehensive comparison between quantile regression methods and traditional geostatistical models for modelling high temperatures remains unexplored.

In this context, the present study aims to compare three different spatiotemporal approaches to estimate daily maximum temperatures at high spatiotemporal resolution: a benchmark satellite-based method (ERA5-Land), and two station-based approaches that can account for covariates (e.g. altitude) or are tailored to model high quantiles of the temperature distribution. We then assess how the differences in the estimated temperatures using these three methods lead to differences in the estimated temperature-mortality associations. 

The first approach is based on the Bayesian spatial quantile autoregressive model by \citet{castillo2023spatial}, which models different parts of the temperature distribution, with a particular focus on high quantiles.
The second approach adopts a Bayesian geostatistical framework, widely applied in environmental exposure modelling such as air pollution studies \citep{cameletti2013spatio}. 
Both of these approaches rely on ground-based monitoring stations as input data. 
The third approach, by contrast, uses GCD from ERA5-Land, a satellite-based product commonly employed in epidemiological practice due to its accessibility and broad temporal coverage \citep{ballester2023heat, ceccarelli2024understanding}. 

Most literature on the temperature-mortality association uses aggregated data at the city or regional level \citep{barnett2007temperature, roye2020comparison}, whilst a limited number of studies rely on individual-level information. In both cases, the temperature-mortality association is known to be non-linear, usually modelled with splines \citep{armstrong2011association}, to obtain what are often described as U-, V-, or J-shaped relationships \citep{braga2002effect}. Use of individual data allows investigation of possible effect modification by individual factors, and it avoids ecological bias \citep{shafran2017ecological}. We took advantage of the individual data availability of the outcome and adopted a case-crossover study design, where each case serves as its own control, thus accounting for time-invariant variables like socioeconomic characteristics \citep{maclure1991case}.

The contributions of this work can be divided into two stages: 
\begin{enumerate}
    \item Spatiotemporal modelling of daily maximum temperatures to estimate an exposure surface to high temperatures;
    \item Epidemiological modelling to assess the association between high temperatures and cause-specific mortality.
\end{enumerate}

As a case study, we use the Lazio region in Italy for the period covering 2008 to 2022 at the municipality level. Analyses are restricted to the summer months, the warmer part of the year. Italy is one of the European countries most vulnerable to the effects of climate change, with projections indicating a marked increase in the frequency, duration, and intensity of extreme heat events in the coming decades \citep{de2019future}. Within this national context, the Lazio region, located in central Italy, offers a particularly interesting case study due to its diverse geography and population distribution. The region includes mountainous areas in the Apennines, extensive coastal zones along the Tyrrhenian Sea and several lakes, for a total land area of 17,242 km$^2$. The capital city of the country, Rome, is located in the region, and this offers the possibility to study a mix of densely populated urban centers and more rural areas. The region is home to 5,714,882 inhabitants, 2,746,984 of which live in the municipality of Rome.

The rest of the paper is structured as follows. Section \ref{sec:methods} presents the data sources used, the three methods for estimating the temperature surface, and the epidemiological model. Section \ref{sec:results} describes the models' estimates, and Section \ref{sec:conclusions} ends the paper with the conclusions and future work. Supplementary Materials accompanying this paper appear online.

\section{Data and methods}\label{sec:methods}

\subsection{Data sources}


All analyses are restricted to the Lazio region of Italy; see Figure S1 of the Supplementary Materials.
The study period spans from 1-1-2008 to 31-12-2022, reflecting the availability of consistent monitoring data; the analyses are restricted to the warm months: May to September in the first stage, to capture the broader high-temperature season, and June to August in the second stage, to focus specifically on the summer period.

In the first stage of the analysis, we used two sources of temperature data: (a) the daily maximum temperatures recorded by 222 monitoring stations in the region, provided by the Italian Institute for Environmental Protection and Research (ISPRA); (b) maximum daily temperatures computed from the hourly air temperature at 2 meters above sea level, extracted from the ERA5-Land dataset \citep{munoz2021era5}. In addition, we considered the average municipality altitude provided by the Italian National Institute of Statistics (ISTAT).

In the second stage of the analysis, we extracted cause-specific individual mortality records from the ISTAT death certificate archives in the time period under study. We further selected only adult deaths (18+ years) with the underline cause due to diseases of the circulatory system and respiratory system, I00-I99 and J00-J99, based on the International Classification of Diseases (ICD-10) codes \citep{world2016world}.

\subsection{Stage 1: Spatiotemporal model of maximum daily temperature}\label{section:spatiotemporal}
In this section, we compare three different approaches for estimating municipality-level daily maximum temperatures in the Lazio region, denoted by ${D} \subset \mathbb{R}^2$. 
For all three methods, we aim to derive point estimates of temperature that will be used as covariates in the second-stage epidemiological models.

\subsubsection{Method 1: Geospatial quantile regression model}
\paragraph{\bf{Spatial quantile autoregression}} 
This approach is based on the Bayesian Geospatial  quantile regression model (GQRM) proposed by \cite{castillo2023spatial}, for quantile levels $\tau \in T = \{0.05, 0.10, 0.20, \ldots, 0.80, 0.90, 0.95 \}$. Let $Y_{t\ell}(\mathbf{s})$ denote the daily maximum temperature of day $\ell$, year $t$, and site $\mathbf{s} \in {D}$. For each quantile level $\tau$, we considered the following model: 
\begin{align} \label{eq:Q}
    Y_{t\ell}(\mathbf{s}) & = Q_{Y_{t\ell}(\mathbf{s})}(\tau\mid Y_{t,\ell-1}(\mathbf{s}))+\epsilon_{t\ell}^\tau(\mathbf{s})  \nonumber \\ 
    & = q_{t\ell}^\tau(\mathbf{s})+\rho^\tau(\mathbf{s})\left(Y_{t,\ell-1}(\mathbf{s})-q_{t,\ell-1}^\tau(\mathbf{s})\right)+\epsilon_{t\ell}^\tau(\mathbf{s})
\end{align}
where $Q_{Y_{t\ell}(\mathbf{s})}(\tau\mid Y_{t,\ell-1}(\mathbf{s}))$ is the $\tau$-level conditional quantile of $Y_{t\ell}(\mathbf{s})$ given $Y_{t,\ell-1}(\mathbf{s})$, and the error term is assumed to follow an asymmetric Laplace (AL) distribution, as is common in Bayesian quantile regression \citep{yu2001}; specifically, $\epsilon_{t\ell}^\tau(\mathbf{s})\sim \text{ind. AL}(0, \sigma^\tau(\mathbf{s}),\tau)$. 
The spatially varying error scale $\sigma^\tau(\mathbf{s})$ and autoregression $\rho^\tau(\mathbf{s})$ parameters are modelled via spatial Gaussian processes on transformed scales. 
In particular, $\sigma^\tau(\mathbf{s})$ 
captures spatial scale dependence through the Gaussian process $Z_\sigma^\tau(\mathbf{s})=\log\{\sigma^\tau(\mathbf{s})\}$, with mean $Z_\sigma^\tau$ and exponential covariance function having variance parameter $\sigma_\sigma^{2,\tau}$ and decay parameter $\phi_\sigma^\tau$. In addition, $\rho^\tau(\mathbf{s})$ 
captures spatial autoregression dependence through the Gaussian process $Z_\rho^\tau(\mathbf{s})=\log\{(1+\rho^\tau(\mathbf{s}))/(1-\rho^\tau(\mathbf{s}))\}$, with mean $Z_\rho^\tau$ and exponential covariance function having variance parameter $\sigma_\rho^{2,\tau}$ and decay parameter $\phi_\rho^\tau$. 

The term $q_{t\ell}^\tau(\mathbf{s})$ contains fixed and random effects as: 
\begin{equation}
q_{t\ell}^\tau(\mathbf{s})=\beta_0^{\tau}+\alpha^\tau t+\beta_1^\tau\sin(2\pi\ell/365)+\beta_2^{\tau}\cos(2\pi\ell/365)+\beta_3^{\tau}X(\mathbf{s})+\gamma_t^\tau(\mathbf{s})
\end{equation} 
Here, $\beta_0^{\tau}$ denotes a global intercept, $\alpha^\tau t$ represents a baseline long-term linear trend in years,  the harmonic terms capture the seasonal behaviour, and $X(\mathbf{s})$ is the standardised altitude at $\mathbf{s}$. Remaining spatiotemporal dependence is captured by the random effects $\gamma_t^\tau(\mathbf{s}) = \beta_0^\tau(\mathbf{s})+\alpha^\tau(\mathbf{s})t+\psi_t^\tau+\eta_t^\tau(\mathbf{s})$. In particular, $\beta_0^\tau(\mathbf{s})$ is a zero-mean Gaussian process with exponential covariance function having variance parameter $\sigma^{2,\tau}_{\beta_0}$ and decay parameter $\phi^\tau_{\beta_0}$, and denotes a local spatial intercept providing location-specific adjustments to the global intercept. Similarly, $\alpha^\tau(\mathbf{s})$ is a zero-mean Gaussian process with exponential covariance function having variance parameter $\sigma_\alpha^{2,\tau}$ and decay parameter $\phi^\tau_\alpha$, and represents a local spatial long-term trend coefficient. 
Lastly, $\psi_t^\tau \sim \text{i.i.d. } N(0,\sigma_\psi^{2,\tau})$ and $\eta_t^\tau(\mathbf{s})\sim \text{i.i.d. } N(0,\sigma_\eta^{2,\tau})$ represent global and local annual intercepts, respectively. Both terms are modelled as independent Gaussian random variables and account for year-to-year variability, such as anomalously hot or cold years.


Model inference is conducted within a Bayesian framework \citep[see Section S1 of the Supplementary Materials or][for the full details]{castillo2023spatial}. Conditionally conjugate prior distributions for all model parameters were specified whenever available to complete the model specification. To perform posterior inference, a Metropolis-within-Gibbs algorithm was employed to obtain MCMC samples from the joint posterior distribution. 

\begin{figure}
    \centering
    \includegraphics[width=1\linewidth]{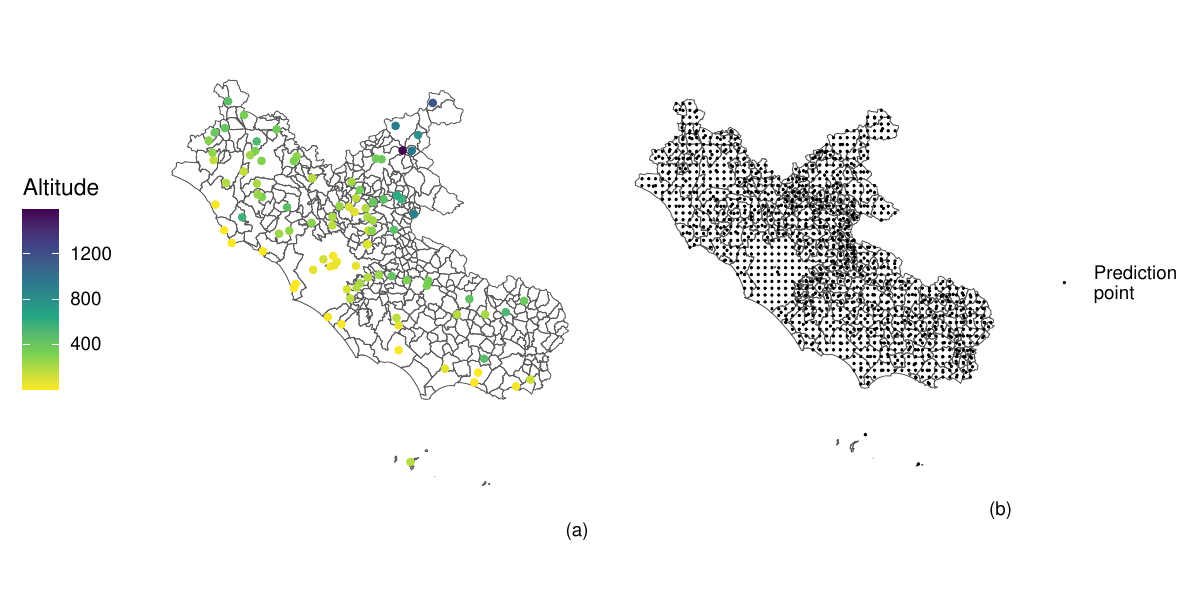}
    \caption{Monitoring stations with altitude information (a), and prediction grid $G_1$ (b) in the Lazio region.}
    \label{fig:mapA}
\end{figure}

As the analysis period spans from May 1st to September 30th, $\ell=2,\ldots,153$ (first day is not modelled in the autoregression), while the index for the year $t=1,\ldots,15$, given that we consider 2008 to 2022. We selected $n_1=93$ monitoring sites, denoted by $S_1$, that satisfy the condition of having no more than seven consecutive days of missing data. The remaining missing values were imputed using splines, as the AL distribution, although suitable in our case as a working likelihood for consistent quantile estimation \citep{sriram2013}, does not represent the true data likelihood and should be used carefully to impute the previous day's temperature.
The reduction in the number of monitoring stations was due to the high proportion of missing data in the original dataset, as shown in Figure S2 of the Supplementary Materials. A detailed data selection flowchart illustrating this process is provided in the Supplementary Materials (see Figure S3).
The locations of the monitoring sites, and their respective altitude, are shown in Figure \ref{fig:mapA}(a).

\paragraph{\bf{Thin plate spline regression}} 
After fitting the GQRMs, we plugged the posterior median of the model parameters in Equation \ref{eq:Q} to obtain the conditional quantile estimates $Q^*_{Y_{t\ell}(\mathbf{s}_i)}(\tau\mid Y_{t,\ell-1}(\mathbf{s}_i))$ for each $\mathbf{s}_i \in S_1$, $t=1,\ldots,15$, $\ell=2,\ldots,153$, and $\tau \in T$.

To construct a spatial exposure map for each day, we need the conditional quantile $Q_{Y_{t\ell}(\mathbf{s}_0)}(\tau\mid Y_{t,\ell-1}(\mathbf{s}_0))$ at arbitrary locations $\mathbf{s}_0\in{D}$. However, the lagged temperature $Y_{t,\ell-1}(\mathbf{s}_0)$ is not observed at unmonitored locations. To approximate the exposure surface, we define a spatial grid $G_1$, which combines the 378 centroids of municipalities with 1000 equally spaced points across the region, resulting in a total of $\lvert G_1 \rvert = 378 + 1000 = 1378$ points that densely and evenly cover the region ${D}$. The locations of the grid are shown in Figure \ref{fig:mapA}(b). 

For each year $t$ and day $\ell$, we interpolate the estimated conditional quantiles across space employing a simple thin plate spline (tps) regression \citep{wood2003thin}. Specifically, we fit the model: \begin{equation}
    Q^*_{Y_{t\ell}(\mathbf{s})}(\tau\mid Y_{t,\ell-1}(\mathbf{s})) = f_{t\ell}(\mathbf{s}) + \epsilon_{t\ell}(\mathbf{s})
\end{equation}
where $f_{t\ell}(\cdot)$ is a smooth spatial function estimated by $\hat{f}_{t\ell}(\cdot)$ minimizing the penalized sum of squares: \begin{equation}
    \sum_{i=1}^{n_1}\left(Q^*_{Y_{t\ell}(\mathbf{s}_i)}(\tau\mid Y_{t,\ell-1}(\mathbf{s}_i)) - f_{t\ell}(\mathbf{s}_i) \right)^2 + \lambda J(f_{t\ell})
\end{equation}
where $J(\cdot)$ is a penalty functional measuring the wiggliness of $f_{t\ell}$ and $\lambda$ controls the trade-off between data fitting and smoothness of $f_{t\ell}$. We set $\lambda = 0.001$ to ensure that the estimated surface closely follows the estimated quantiles. 

The interpolated exposure surface across the region is then defined as $\hat{Q}^*_{Y_{t\ell}(\mathbf{s}_j)}(\tau \mid Y_{t,\ell-1}(\mathbf{s}_j)) = \hat{f}_{t\ell}(\mathbf{s}_j)$ for each $\mathbf{s}_j \in G_1$.  Municipality-level exposures are finally obtained by averaging the interpolated values across all grid points $\mathbf{s}_j$ falling within each municipality.

\subsubsection{Method 2: Geospatial Gaussian process model}
The second approach is a Geospatial Gaussian process model (GGPM) that describes the conditional mean of the response variable. Therefore, results can be expected to be comparable to those obtained with the quantile regression at $\tau=0.5$. 

Following the notation above, the model can be written as:
\begin{equation}
    Y_{t\ell}(\mathbf{s})=\beta_0+\beta_1 X(\mathbf{s})+\xi_{t\ell}(\mathbf{s})+\epsilon_{t\ell}(\mathbf{s})
\end{equation}
In addition to the intercept $\beta_0$ and the standardised altitude covariate $X(\mathbf{s})$, the model includes a measurement error term $\epsilon_{t\ell}(\mathbf{s}) \sim \text{i.i.d. N}(0, \sigma_\epsilon^2)$, modelled as a Gaussian white-noise process, and a state process $\xi_{t\ell}(\mathbf{s})$, which represents the latent spatiotemporal signal. Specifically, the spatial process $\xi_{t\ell}(\mathbf{s})$ evolves over time according to a first-order autoregressive process \citep{harvill2010spatio}, and is specified as: \begin{equation}
    \xi_{t\ell}(\mathbf{s}) = a \xi_{t,\ell-1}(\mathbf{s})+\omega_{t\ell}(\mathbf{s}) 
\end{equation}
for $\ell=2,\ldots,153$ and coefficient $\lvert a \rvert < 1$. For $\ell=1$, the process is initialized from its stationary distribution \begin{equation}
    \xi_{t1}(\mathbf{s})\sim \text{i.i.d. }\text{N}\left(0, \frac{\sigma_\xi^2}{1-a^2}\right)
\end{equation}
Moreover, $\omega_{t\ell}(\mathbf{s})$ is a zero-mean Gaussian field, independent in time and, for the same time point, have a variance $\sigma^2_\omega$ and a correlation function $\mathcal{C}(h)$. Here, the correlation function $\mathcal{C}(h)$ is purely spatial and it is a Matérn function: \begin{equation}
    \mathcal{C}(h) = \frac{1}{\Gamma(\nu)2^{\nu-1}}(kh)^\nu K_{\nu}(kh)
\end{equation}
where $h = \lVert \mathbf{s} - \mathbf{s}^{\prime}\rVert$ is the absolute Euclidean distance between two spatial points $\mathbf{s}, \mathbf{s}^{\prime}\in {D}$, $K_\nu$ denoting the modified Bessel function of the second kind and order $\nu>0$. The parameter $\nu$ measures the degree of smoothness of the process, and the scaling parameter $k > 0$ controls the spatial correlation range.

\begin{figure}
    \centering
    \includegraphics[width=1\linewidth]{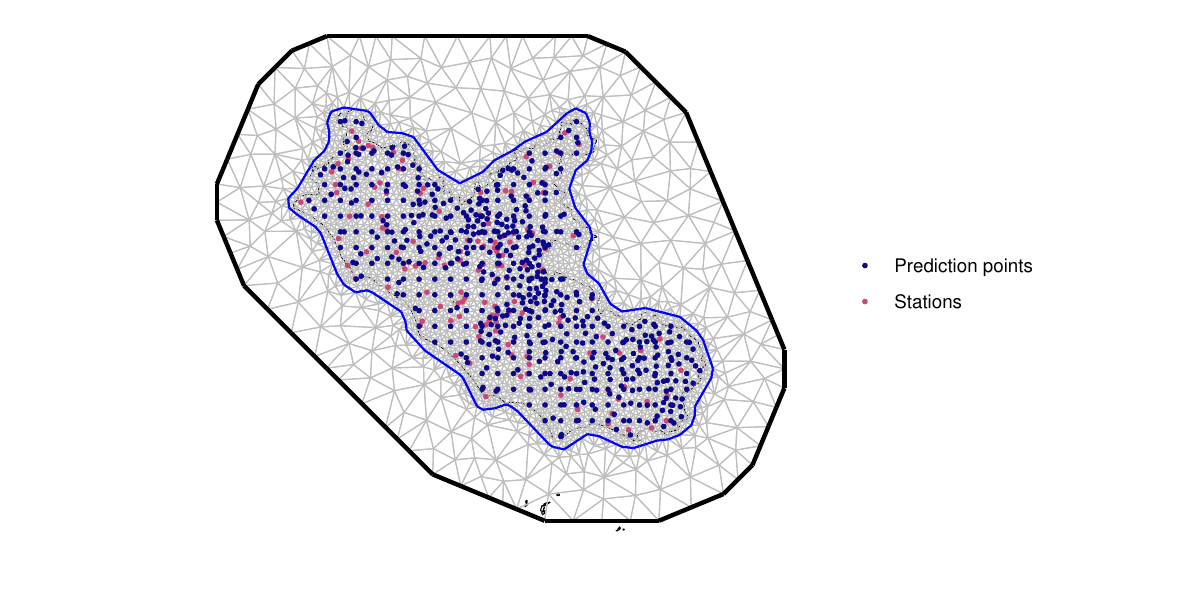}
    \caption{Plot of the mesh in use with the location of the stations $S_2$ and the prediction points $G_2$.}
    \label{fig:mesh}
\end{figure}

The model is implemented within the  \texttt{R-INLA} package \citep{rue2009approximate, martins2013bayesian, van2023new}.
using the SPDE approach \citep{lindgren2011explicit, blangiardo2013spatial, cameletti2013spatio, riley2025bayesian} to represent a continuous spatial process as a discretely indexed spatial random process. This produces substantial computational advantages since it avoids the so-called \textit{big n problem} that affects spatial Gaussian processes \citep{jona2013discussing}.  
As shown in \cite{lindgren2011explicit}, within the SPDE approach the spatial correlation structure is modelled by solving the linear fractional SPDE: \begin{equation}
    (k^2-\Delta)^{\alpha/2}(\tau u(\mathbf{s}))=W(\mathbf{s})
\end{equation}
with $k>0$ spatial scale parameter, $\alpha$ governs the smoothness of the field, $\Delta$ is the Laplacian operator, and $W(\mathbf{s})$ is Gaussian white noise. This yields a Matérn covariance function with smoothness $\nu = \alpha - d/2$, where $d$ is the spatial dimension (here $d = 2$). The approximation employs a linear combination of basis functions, defined in a triangulated domain called a mesh.
In our application, to discretize the spatial domain, we construct a non-convex mesh using a resolution adapted to the density of prediction and observation points. We constructed the mesh shown in Figure \ref{fig:mesh}, with a fine maximum edge length of 4.5~km and 20~km, for the inner and outer meshes, respectively, and the boundaries determined by the full grid coordinates.

To fit the model, we selected $n_2=140$ monitoring sites, denoted by $S_2$, under the constraint that each site has less than $20\%$ missing observations (see Supplementary Figure S2). This approach leverages the ability of the Bayesian approach to handle missing data through model-based imputation. Due to computational limitations, we estimated a separate model for each year $t = 1,\ldots,15$, treating them independently.  Given the observed values of $Y_{t\ell}(\mathbf{s})$, we rely on the posterior predictive distribution to obtain realisations $Y_{t\ell}(\mathbf{s}_j)$, for each grid point $\mathbf{s}_j\in G_2$, built as before but with 250 equally spaced points for a total of $\lvert G_2 \rvert=628$ due to computationally restrains, $\ell=1,\ldots,153$ and $t=1,\ldots,15$. In the prediction stage, each location $\mathbf{s}_j\in G_2$ was assigned the average altitude of the municipality in which it is located. The Bayesian approach, through the INLA algorithm, provides the posterior distribution of $\xi$ for all the $n_2+\lvert G_2 \rvert$ triangulation vertices, so it is straightforward and computationally convenient to obtain predictions for $Y_{t\ell}(\mathbf{s}_j)$ as: \begin{equation}
    f(Y_{t\ell}(\mathbf{s}_j)\mid \mathbf{y})=\int f(Y_{t\ell}(\mathbf{s}_j)\mid \bm{\theta}) f(\bm{\theta}\mid \mathbf{y}) \, \text{d}\bm{\theta}
\end{equation} 
where $\bm{\theta}$ is the set of model parameters and $\mathbf{y}$ are the observed data.
Lastly, as in the previous method, we can compute municipality-level temperatures as average of all $\mathbf{s}_j$ locations falling within each municipality.

\subsubsection{Method 3: ERA5-Land data}
As a benchmark, we consider the method commonly used in the literature, which is based on hourly air temperatures extracted from the ERA5-Land dataset \citep{munoz2021era5}. We first computed for each day the maximum temperature and then obtained municipality-level values using area-weighted averages. Given $\{C_1, \ldots, C_K\}$, the ERA5-Land grids that fall within the Lazio region, the temperature for the municipality $m_i$ on day $\ell$ and year $t$ is given by: \begin{equation}
    Y_{t\ell}(m_i) = \frac{1}{A_{m_i}}\sum_{k=1}^K A_{m_ik} Y_{t\ell}(C_k)
\end{equation} 
where $A_{m_i}$ is the total surface of the municipality $m_i$, for $i=1,\ldots,378$, and $A_{m_ik}$ is the surface of the cell $C_k$ that falls in the municipality $m_i$. The total surface of the municipality can also be expressed as $A_{m_i} = \sum_{k=1}^KA_{m_ik}$, this is the sum of the proportion of the grid cells within its borders.

\subsection{Stage 2: Epidemiological modelling}
In the second stage of the analysis, we assessed the association between exposure to high temperatures and mortality due to cardiovascular and respiratory causes. To this end, we used the temperature estimates obtained from the three spatiotemporal methods described in the previous section as exposure in the following epidemiological models.

We adopted a time-stratified case-crossover design, commonly used for analysing the effect of transient exposures \citep{lu2007equivalence, navidi2002risk}. The temperature on the day of the cardiovascular/respiratory death (event day) is compared with the temperature on non-event days. Within this design, each case acts as its own control, intrinsically adjusting for individual-level factors that do not vary over time (e.g. age, sex, ethnicity, deprivation). We selected non-event days as the same day of the week, calendar month and year as the event day to avoid overlap bias \citep{konstantinoudis2022ambient, konstantinoudis2023asthma}. 
Following \cite{deplazio2004heatwaves, michelozzi2009high} and \cite{singh2024heat}, each individual case was linked to the average daily maximum temperature in the 3 days before the event day and the corresponding control days.

In order to simplify notation, differently from the spatiotemporal exposure models in Section \ref{section:spatiotemporal}, here we use $t$ to denote days between June and August in the years 2008 to 2022.

Let $O_{tj}$ be the case-control identifier at time $t$ and in the $j$-th case-control group. Then $X_{tjh}$ is the exposure for the $j$-th case-control group, defined as the mean daily maximum temperature estimated by the $h$-th method ($h =1,2,3$ as in Section \ref{section:spatiotemporal}), in the 3 days prior to the event (or control) day $t$ Finally, $\text{holiday}_t$ is a binary variable 1/0 indicating if the $t$-th day is a bank holiday. We specified Bayesian hierarchical conditional Poisson models as:
\begin{align} \label{model_epi1}
&O_{tj} \sim \text{Poisson}(\mu_{tj}) \nonumber\\
&\log(\mu_{tj}) = \beta_0 + f_{\text{rw2}}(X_{tjh}) + \beta_1 \cdot \text{holiday}_t + u_j \nonumber\\
&X_{tjh}\mid X_{(t-1)jh},X_{(t-2)jh},\tau_X\sim \text{N}(2X_{(t-1)jh}-X_{(t-2)jh},\tau_X^{-1})\nonumber\\
&\tau_X \sim \text{PCprior}\left( P\left( \tau_X^{-1/2} > 0.1 \right) = 0.01 \right)\nonumber\\
&u_j \sim \text{N}(0, 100)\nonumber\\
&\beta_0, \beta_1 \sim \text{N}(0, 1000)
\end{align}
where $\beta_0$ and $\beta_1$ are the coefficients associated with the intercept and the holiday variable, respectively; $f_{\text{rw2}}(\cdot)$ is a second-order random walk, used to model the temperature effect; and $u_j$ is a fixed effect corresponding to the event/non-event day grouping. We adopted weakly informative priors for the regression coefficients and the group effect, while a penalised complexity prior was specified for the second-order random walk smoothing term to ensure interpretable regularisation and to avoid excessive flexibility unsupported by the data. 

\subsubsection{Heatwaves}\label{sec:heatwaves}
The definition of a heatwave is not straightforward, as it involves both a temperature metric, a temperature threshold, and the duration of consecutive days exceeding that threshold. In the analyses, we considered four different thresholds and three duration criteria \citep{xu2016impact, guo2017heat, kang2020heatwave} for daily maximum temperatures. For the threshold we evaluated three empirical quantiles ($0.90$, $0.925$, $0.95$) of the daily maximum temperature time series in the summer for each municipality, as well as a fixed value at 35$^\circ$C \citep{vecellio2022evaluating, vanos2023physiological}. For the duration we considered: (a) surpassing the threshold, which is labelled as \textit{heatwave\_base}; (b) being exposed to at least 2 consecutive days above the threshold; (c) being exposed to at least 3 consecutive days above the threshold. For (b) and (c) we also excluded the first day that the temperature surpasses the threshold as a heatwave day, to better capture the prolonged effect of sustained heat \citep{rocklov2012estimation}. These two scenarios are labelled as \textit{heatwave\_1daylag} and \textit{heatwave\_2dayslag}, respectively. Similarly to the temperature lag, a subject is considered affected by a heatwave if a heatwave day occurs in their municipality on the day of the event or during any of the three preceding days.

We estimated models including the heatwave variable, using the same specification as in Equation \ref{model_epi1}, with the addition of a binary covariate $\text{heatwave}_{tjhz}$. The variable indicate whether, in the three days preceding the event day $t$, the $j$-th group was exposed to at least one heatwave day, defined according to the $h$-th spatiotemporal model and the $z$-th combination of temperature threshold and heatwave definition. The corresponding regression coefficient, $\beta_2$, was assigned a prior distribution $\beta_2 \sim \text{N}(0, 1000)$, consistent with the other fixed effects.


\subsection{Implementation}
All analyses were conducted using \texttt{R} \citep[version~4.4.2,][]{Rstudio}. The GQRM was fitted using the MCMC algorithm described in \cite{castillo2023spatial}, while tps's were estimated using the \texttt{fields} package \citep{fields_package}. GGPM and epidemiological models were implemented with \texttt{R-INLA} \citep{Lindgren_INLA2015}. Municipality-level area weighted averages from the ERA5-Land data were computed using the \texttt{extract} function of the \texttt{terra} package \citep{terra2025}.

\section{Results}\label{sec:results}

\subsection{Spatiotemporal models}
In the following, we compare the temperatures obtained using the three different methods presented in Section \ref{section:spatiotemporal}. For method 1, GQRM, we show results on the median ($\tau = 0.5$) in order to allow comparison with the other two approaches.

Prior to presenting the modelling results, Figure S4 in the Supplementary Materials shows a quantile--quantile (QQ) plot comparing daily maximum temperatures from ground monitoring stations with the nearest ERA5-Land reanalysis grid cell within the study region. The plot indicates that while ERA5-Land tends to slightly overestimate lower temperature values, it systematically underestimates those in the upper range.

\begin{figure}[tbh]
    \centering
    \includegraphics[width=1\linewidth]{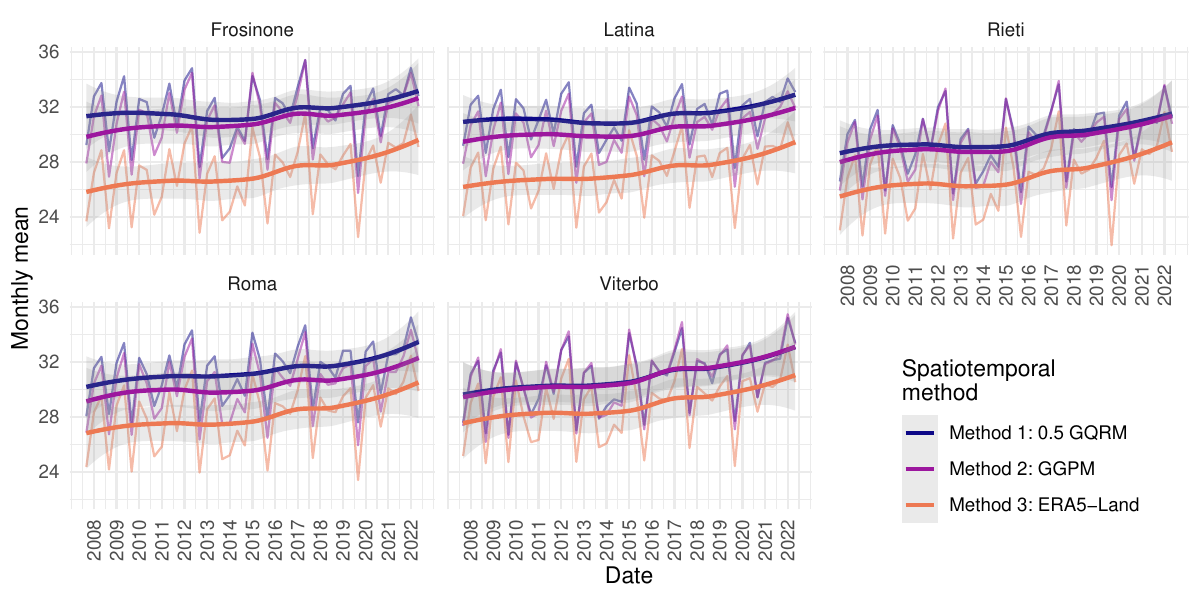}
    \caption{Averages of daily maximum temperatures ($^\circ$C) in the summer months, estimated by the three methods across the five provinces of Lazio. The general trend is drawn using LOESS regression with a $95\%$ confidence interval.}
    \label{fig:trend_prov}
\end{figure}

Figure \ref{fig:trend_prov} displays the temporal trend of estimated daily maximum temperatures produced by the three spatiotemporal methods. The difference between the station-based approaches and the Copernicus ERA5-Land reanalysis is immediately evident, particularly in the warmest provinces of the region, Frosinone and Latina. An overall increasing trend in temperature over the years is also observed.

Yearly estimates between the three methods are shown in Table S1 of the Supplementary Materials, and it is possible to notice the growing overall trend and how the differences between the methods vary slightly over the years. 

\begin{figure}
    \centering
    \includegraphics[width=1\linewidth]{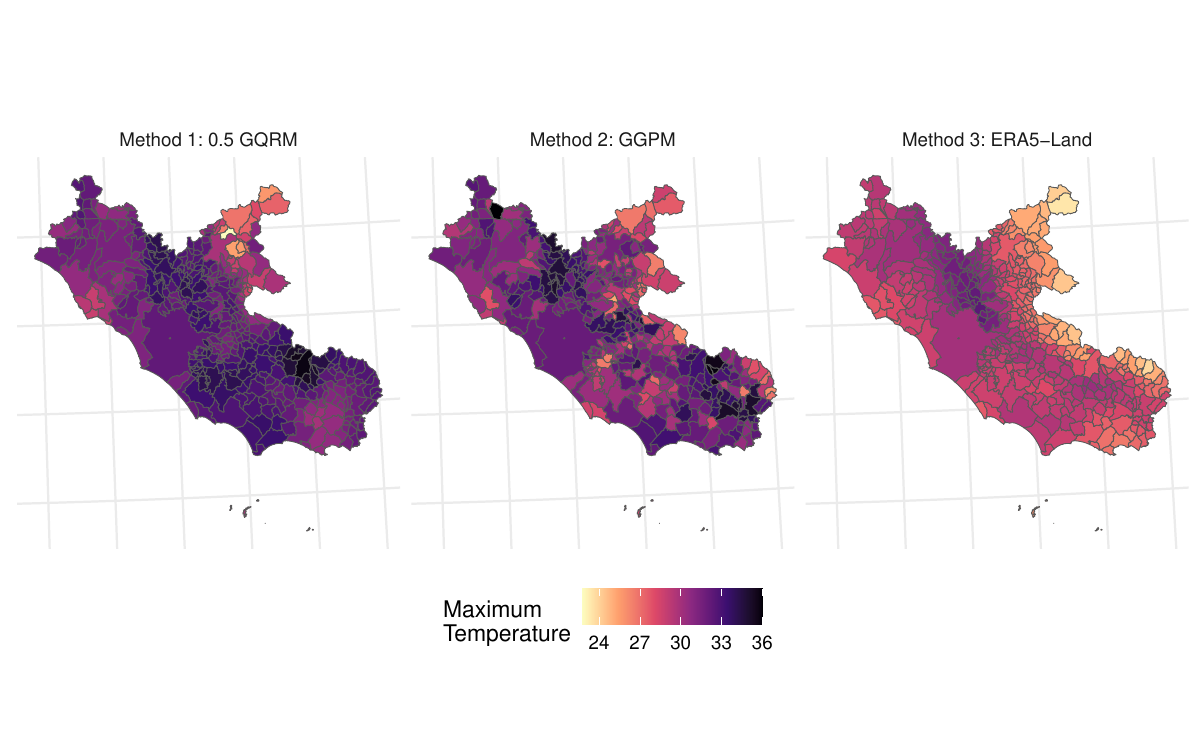}
    \caption{Average daily maximum temperature ($^\circ$C) per municipality and spatiotemporal methods in July 2022.}
    \label{fig:mappa_07_22}
\end{figure}

An example of the spatial differences between the three methods is shown in Figure \ref{fig:mappa_07_22}. All models capture the main spatial patterns: the northern part of the region is consistently colder, while the area just north of Rome appears to be the the warmest. In the ERA5-based method, coastal municipalities exhibit a more temperate climate compared to inland areas \citep{munoz2021era5}, a feature not observed in the station-based methods. The GGPM approach displays the greatest variability between neighbouring municipalities, possibly due to the inclusion of altitude as a covariate in the model predictions. All daily and monthly maps comparing the methods are available through an interactive Shiny App.\footnote{\url{https://emilianoceccarelli.shinyapps.io/temperature_comparison_Lazio/}}

\subsection{Epidemiological models}
\begin{figure}
    \centering
    \includegraphics[width=1\linewidth]{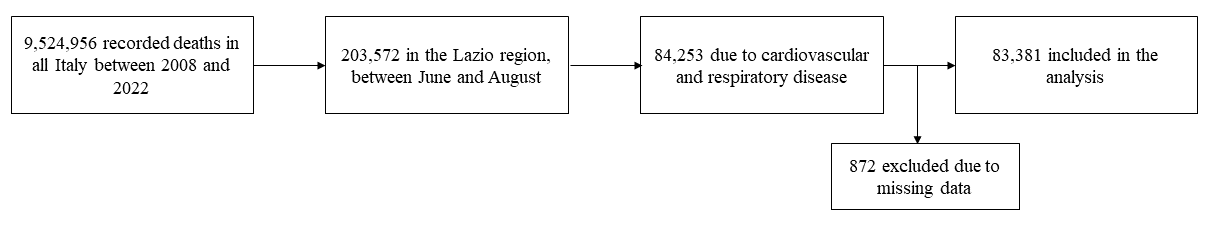}
    \caption{Study profile.}
    \label{fig::study_profile}
\end{figure}

We included 83,381 death events in the study, and a more detailed study profile is reported in Figure \ref{fig::study_profile}. Of the total, $45\%$ were males and $55\%$ females. The average age at death was 83 years (interquartile range: 79--91). The spatial distribution was uneven, with $71\%$ of deaths occurring in the province of Rome, reflecting the underlying population density. After creating the control cases, each model was fitted on a total of 369,901 observations (83,381 cases and 286,520 controls).

\begin{figure}
    \centering
    \includegraphics[width=1\linewidth]{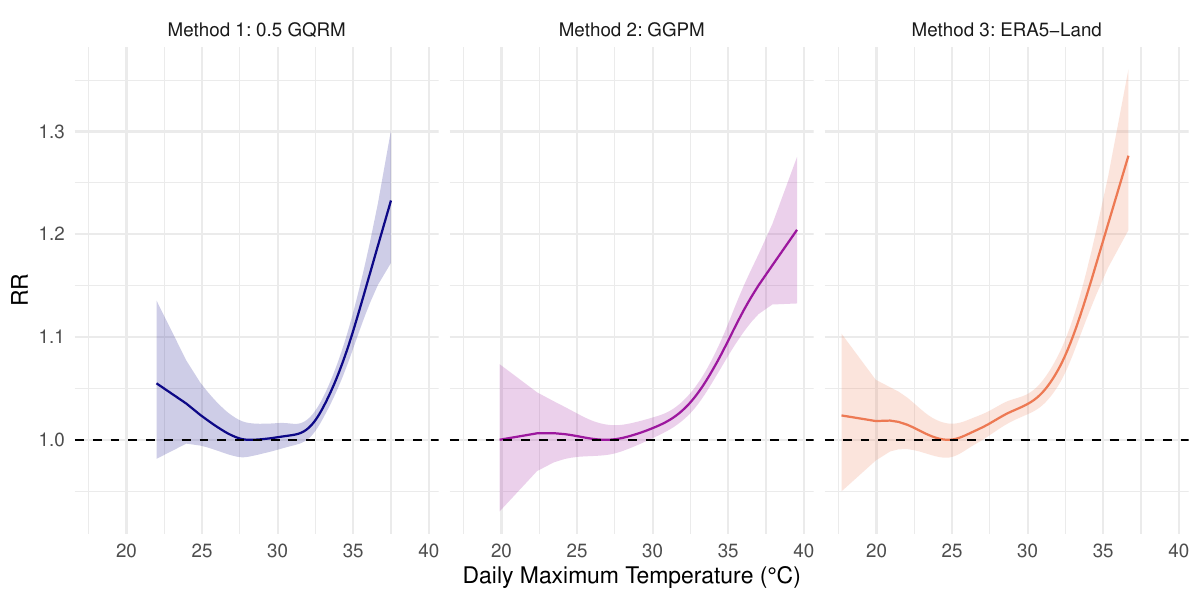}
    \caption{RR estimates for daily maximum temperature exposure by spatiotemporal methods. Shaded area indicates $95\%$ credible interval.}
    \label{fig:RR_temp}
\end{figure}

Figure \ref{fig:RR_temp} shows the relationship between temperature and mortality, expressed in terms of relative risk (RR). We fitted the models described in Equation \ref{model_epi1}, applying temperature exposures derived from the three different spatiotemporal methods. As in the previous section, the figure presents results for the 0.5 GQRM only to facilitate comparison with the other two approaches. Detailed plots for all GQRMs considered in the analysis are available in the Supplementary Materials (see Figure S5). Each curve has been normalised so that $RR=1$ at the temperature corresponding to the lowest mortality risk. This temperature is often referred to as the optimal temperature or minimum mortality temperature (MMT) \citep{gasparrini2015mortality, lopez2021evolution}, facilitating interpretation and comparison across exposure surfaces. 
In line with previous literature \citep{armstrong2006models, gasparrini2015mortality}, the estimated RR curves exhibit a J-shaped relationship, with steeper increases in risk at high temperatures. All spatiotemporal methods capture the elevated risk associated with extreme heat, but differ in the MMT. As shown in Table \ref{tab::minRR}, the MMT varies by up to 2$^\circ$C when comparing the ERA5-Land-based estimates to the GGPM, and by more than 3$^\circ$C compared to the median GQRM. Table S2 in the Supplementary Materials reports the MMTs for all quantile levels, showing that the 0.05 GQRM returns an MMT closest to the ERA5-Land estimate.

Estimates for the \textit{holiday} coefficient are in agreement between models and indicate a $-11.27\%$ reduction in mortality risk compared to non-holiday days (95\% credible interval: $-15.66\%$ to $-6.65\%$). 

\begin{figure}[tbh]
    \centering
    \includegraphics[width=1\linewidth]{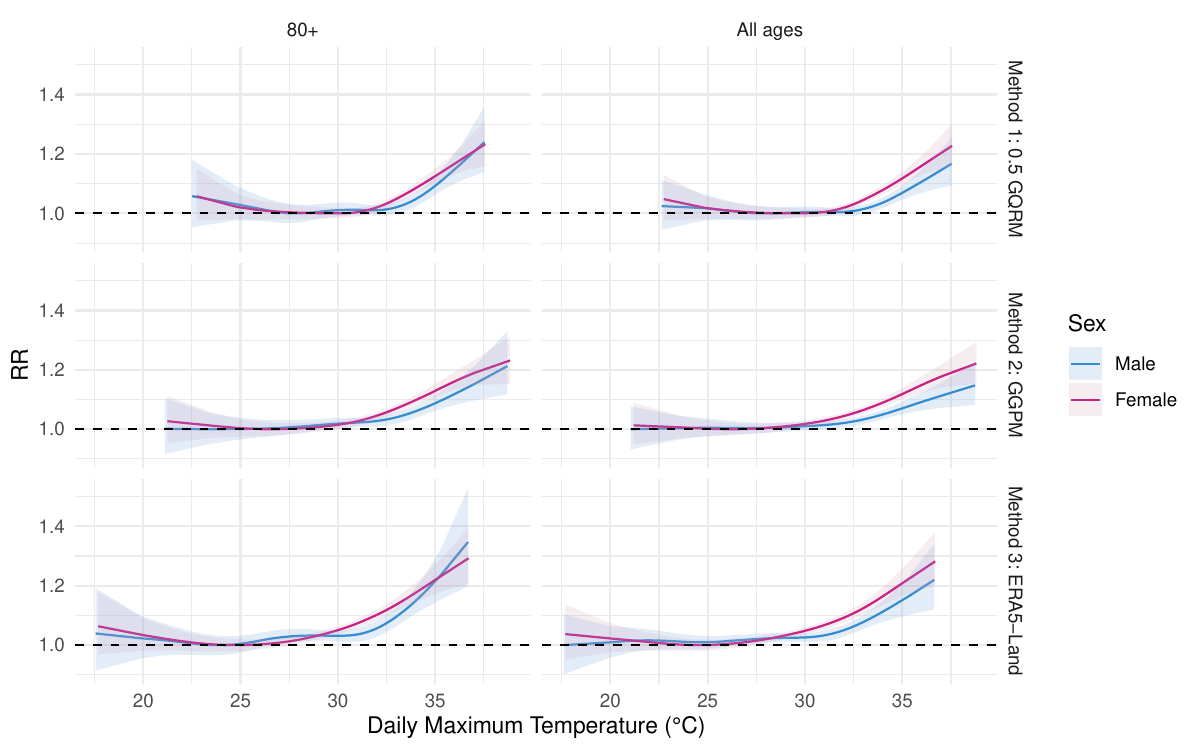}
    \caption{RR estimates for maximum temperature exposure by spatiotemporal method, age group (80+ and total population), and sex. Shaded areas indicates $95\%$ credible intervals.}
    \label{fig:RR_temp_agesex_paper}
\end{figure}

We performed stratified analyses by age class (18--64, 65--79, 80+, and 18+) and sex.  Figure \ref{fig:RR_temp_agesex_paper} shows the RR curves for the sex specific curves across all ages and for the most vulnerable age group (80+). Females show a more consistent increase in risk beginning at lower temperatures, and leading to slightly higher RRs compared to males. 
Additionally, the 80+ group also exhibits a pronounced increase of RR at lower temperatures compared to the total population, indicating high vulnerability to extreme heat.

The complete results are shown in Figure S6 of the Supplementary Materials and overall they are consistent  across spatiotemporal temperature methods. No apparent increase in risk is observed for the 18--64 age group, while a general upward trend is visible for the 65--79 group, especially among females, although there is no strong evidence of a clear difference.

\subsubsection{Heatwaves assessments and model}\label{sec:mod_heatwave}
Each spatiotemporal method and each heatwave definition (as described in Section \ref{sec:heatwaves}) results in a different number of days being formally classified as heatwaves. As shown in Figure S7 of the Supplementary Materials, the differences across temperature datasets are not substantial as long as the threshold is defined based on the municipality-specific time series.

\begin{figure}[tbh]
    \centering
    \includegraphics[width=1\linewidth]{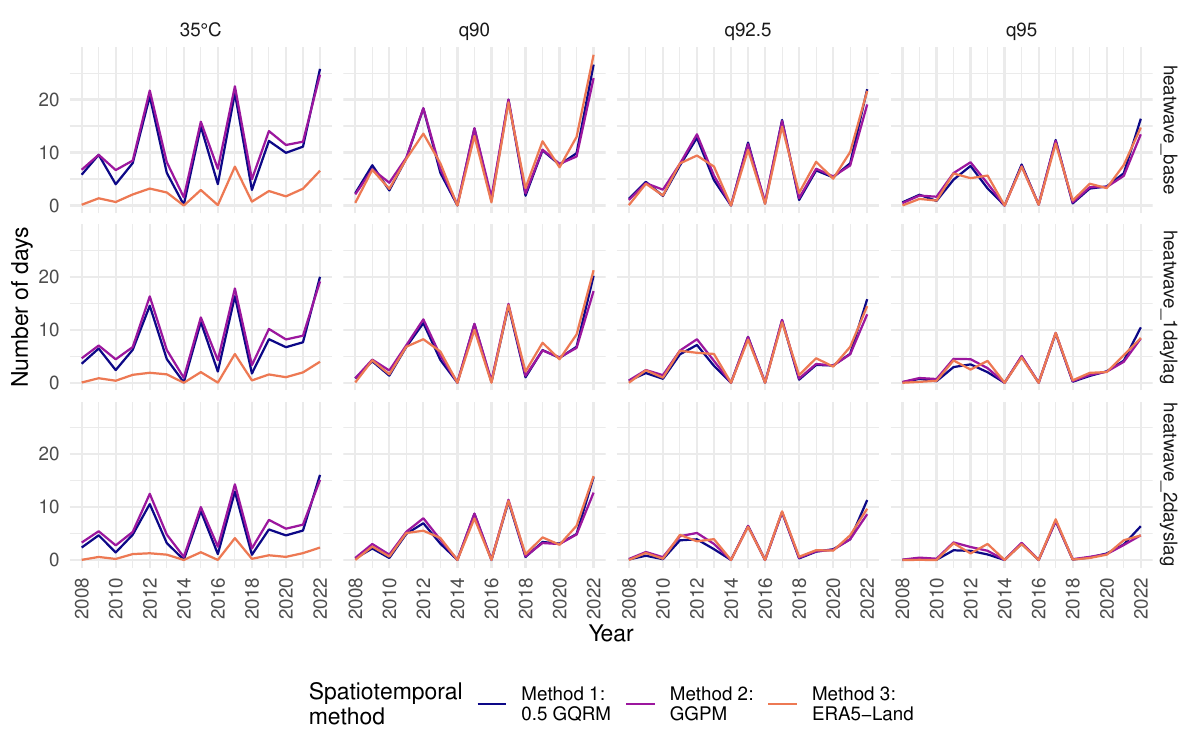}
    \caption{Average annual number of heatwave days per duration criteria, spatiotemporal temperature method and temperature threshold.}
    \label{fig::avg_heatwave}
\end{figure}

Figure \ref{fig::avg_heatwave} presents the average number of days that a municipality in the region has experienced per duration criteria, spatiotemporal temperature method and temperature threshold. The plots highlight the increasing trend in heatwave occurrence over time and shows how the use of a fixed temperature threshold is sensitive to the underlying temperature source. Despite methodological differences, all spatiotemporal models and quantile thresholds consistently capture the temporal dynamics of heatwaves over the study period. 

\begin{figure}[tbh]
    \centering
    \includegraphics[width=1\linewidth]{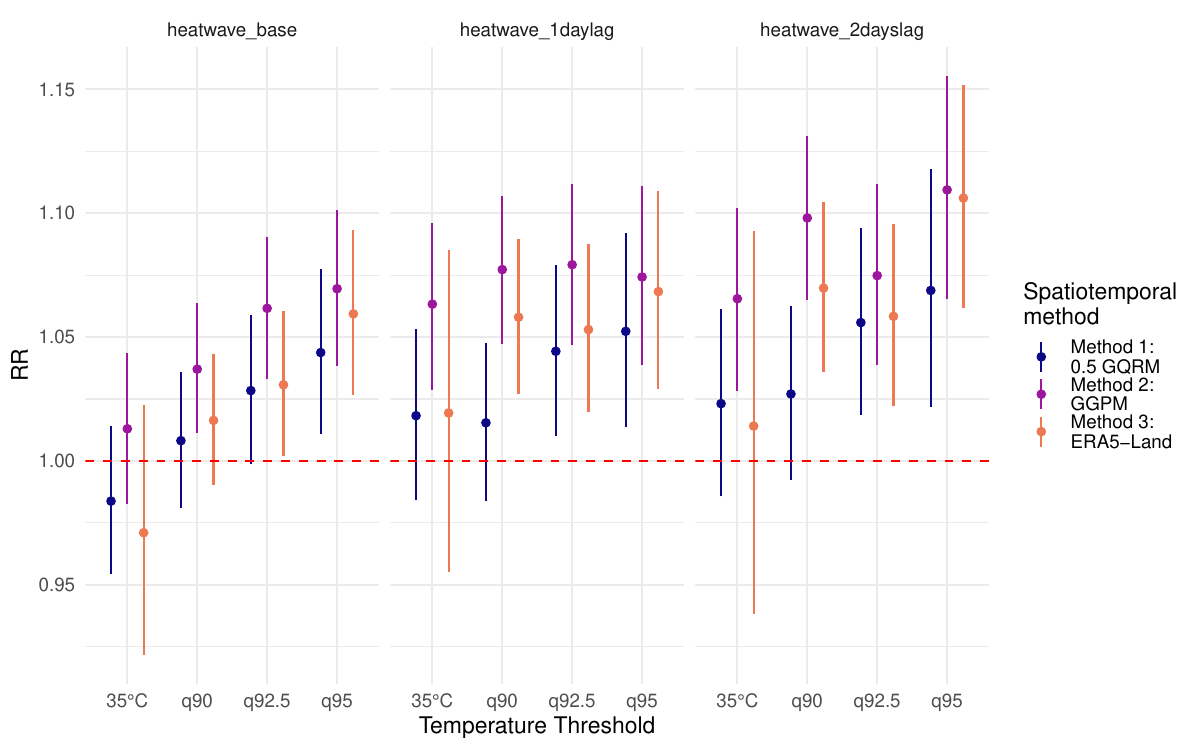}
    \caption{RR estimates and their $95\%$ credible interval of the heatwave variable in the model presented in Section \ref{sec:mod_heatwave}. Results per duration criteria, spatiotemporal temperature method and temperature threshold.}
    \label{fig:heatwave_RR}
\end{figure}

Figure \ref{fig:heatwave_RR} shows the estimated RRs associated with the heatwave variable in the model presented in Section \ref{sec:heatwaves}, along with their corresponding $95\%$ credible intervals. Estimates based on the fixed 35$^\circ$C threshold are quite variable and generally pushed towards 0. This is likely due to multicollinearity with the continuous temperature variable. To support this claim, we estimated again the models without the continuous temperature variable. The results, reported in Figure S8 of the Supplementary Materials, show that the estimated RRs remain consistently above 1 and comparable across spatiotemporal models, duration criteria, and thresholds, even when using the fixed 35$^\circ$C threshold.
Considering the 90th percentile threshold, the relative risks are around 1.02 for the 0.5 GQRM using the three heatwave definitions, while stronger effects are seen for the GGPM and ERA5-Land temperature model, particularly in combination with stricter heatwave duration criteria. As the quantile threshold increases, RR estimates tend to stabilize across spatiotemporal methods; an exception is the 95\% quantile for the two days lag criterion, where the RR estimates are characterised by large uncertainties, suggesting a potential trade-off between threshold severity and statistical power.
The temperature-mortality RR curve retains its overall shape even when the heatwave variable is included, as shown in Figure S9 of the Supplementary Materials. When comparing these curves with those estimated from the previous model (see Figure \ref{fig:RR_temp}), a slight decline is seen in the curves for high temperatures when a balanced heatwave definition is applied, that is, a combination of threshold and duration criteria that captures sustained heat exposure without being overly restrictive. This suggests that part of the effect attributed to high temperatures may be explained by discrete heatwave events.

\subsection{Sensitivity analysis}

As a sensitivity analysis, we estimated models \ref{model_epi1} selecting malignant neoplasm (ICD-10 codes C00--C97) as the underlying cause of death. This outcome was chosen as a negative control because, unlike cardiovascular and respiratory diseases, cancer mortality is not expected to be acutely affected by short-term fluctuations of daily maximum temperature. Additionally, the large number of malignant neoplasm mortality (58,917 cases and 202,086 controls) ensured a sample size comparable to that of cardiovascular and respiratory causes considered in the main analyses.    

Consistent with our expectations, there was not evidence of an association  between high temperatures and mortality, with effect size close to 0 across all spatiotemporal temperature exposure estimation methods. RR estimates are shown in Figure S10 of the Supplementary Materials.

\section{Discussion and conclusions}\label{sec:conclusions}

Using established methods, we compared three approaches for deriving municipality-level exposure surfaces to extreme heat. In the first stage, we estimated daily maximum temperatures using two model-based spatiotemporal methods relying on weather station data, and one GCD reanalysis, Copernicus' ERA5-Land. Our results show spatial variability across the three approaches. ERA5-Land GCD systematically produced lower temperatures in coastal municipalities compared to inland areas, an effect not observed in the station-based approaches. This discrepancy can be explained by the averaging process over mixed land-sea grid cells, which reduces estimates for municipalities bordering the coastline. A similar effect was observed in mountainous regions, where valley settlements may inherit cooler estimates due to the influence of neighbouring high-altitude grid cells. These findings are consistent with previous literature reporting limitations of ERA5-Land in complex terrains, as shown by \cite{mistry2022comparison}, who noted that the ERA5-Land reanalysis might perform worse in tropical regions. The two station-based modelling approaches produced consistently higher temperatures than the satellite-based estimates. Among them, the GQRM combined with thin based splines (Method 1) yielded the highest values. The tps interpolation used in this method provided smoother spatial patterns, whereas the GGPM (Method 2)  captured sharper variability between neighbouring municipalities, reflecting the inclusion of altitude as a covariate. 
We might consider a calibration or data fusion approach in a Bayesian modelling framework. However, in this study, our focus is on detecting the effects of high temperatures; therefore, merging station observations with ERA5-Land estimates could mask potential evidence in this regard.

The three methods were compared in an epidemiological context, assessing the relationship between heat exposure and mortality from cardiovascular and respiratory diseases. In accordance with existing literature, our models found that despite using different temperature sources and methods, the exposure-response J-shaped curve remains largely unchanged. \citet{roye2020comparison} noted that the mortality risk estimates using ERA5-Land were slightly lower than those using measuring stations. This differs from our estimates, as we find a slightly higher RR using the temperatures derived by ERA5-Land, as shown in Figure \ref{fig:RR_temp}. The interpretation of the maximum RR must be considered relative to the MMT, which is lower for ERA5-Land, as reported in Table \ref{tab::minRR}, and thus shifts the point of reference for comparison. A similar result is found by \cite{choi2023heat}, who found more than 3$^\circ$C of difference in MMTs between station-based models and a GCD. Overall, the three approaches provide broadly consistent estimates, reinforcing the robustness of the observed association.

From a public health perspective, the differences observed in the estimated MMTs open a discussion for the design of early warning systems. The wide spatial coverage of ERA5-Land ensures that every municipality can be assigned an exposure value, making it a valuable tool for regional or national surveillance systems. However, its tendency to underestimate temperatures in complex terrains may potentially trigger warnings in advance. Conversely, monitoring stations provide highly accurate observations, but their sparse and uneven distribution limits their applicability, unless coupled with a modelling approach which allows municipality-specific estimates. 

In the Lazio region, the current heat warning system is coordinated by the regional Department of Epidemiology and it is based on four alert levels, with level 3 activated when three consecutive days are forecasted to exceed a municipality-specific threshold of maximum temperature \citep{deplazio}. In this context, differences in MMT estimation across exposure models and temperature sources may directly influence the identification of critical days and the timing of interventions.

Given that early warning systems are inherently based on both intensity and persistence of extreme heat, we further examined the role of heatwaves as prolonged exposure events. Adding the heatwave specification in the models allowed us to investigate whether prolonged exposure to extreme heat conveys an additional risk, beyond that captured by continuous temperature variable. Consistent with previous work \citep{Gasparrini_Armstrong_2011}, we found that including a categorical heatwave indicator did not substantially alter the overall exposure-response curve. However, contrary to \citet{guo2017heat}, who modelled exposure using daily mean temperatures, our results based on daily maximum temperatures showed strong evidence of an for the  heatwave variable after controlling for daily temperature. This suggests that, with an appropriate heatwave definition, epidemiological models can provide further insight into the health risks of prolonged exposure to extreme heat.

When stratifying the analysis by age groups and sex, we found variability in the dose-response curves, particularly for younger ages. As the majority of deaths is observed in the 80+, as expected the dose-response was similar to the all ages one. In both cases, the curves for females tend to rise earlier and remain consistently above those for males. This indicates that  females (and specifically older ones) become at increased risk of heat-related mortality at lower temperatures than their male counterparts. This gendered vulnerability is consistent with the literature on the topic \citep{van2019sex}, but the underlying reasons for this are not well understood. 
Enhanced vulnerability to heat can be explained by a higher prevalence of multi-morbidity in women compared to men, partially because females have a higher life expectancy compared to males \citep{abad2014age}. Many chronic diseases of ageing are exacerbated by high temperatures, or affect thermoregulation directly. For instance, glycaemic regulation is impaired by heat in diabetic patients, but so are other homeostatic activities like sweating \cite{kenny2016body} impairing their ability to dissipate heat. Furthermore, multi-morbidity leads to the use of multiple medications. A recent meta-analysis by \cite{hospers2024effect} identified multiple treatments for ageing-related diseases amongst the ones impairing temperature regulation during heat stress. Although females have been observed to have an inherently impaired ability to thermoregulate, these differences are negligible once body size and surface area have been accounted for \citep{kaciuba2001gender}. On the contrary, it appears that heat illness is much more prevalent in males than in females, \citep{gifford2019risk} which might explain the small increase in heat-related mortality rates in the males youngest age group, despite not being characterised by strong evidence of an effect. 

Thanks to the availability of individual data, we adopted a case-crossover methodology that offers important advantages in environmental epidemiology. By comparing each subject to themselves at different times, it provides strong control for all individual characteristics that do not vary over time, reducing the risk of confounding \citep{janes2005case}. 

Whilst the spatiotemporal approaches allowed us to construct precise exposure profiles within the cities, we were forced to match the aggregation of health data, aligning at the municipality scale. In Lazio, this issue mainly concerns the city of Rome, the only municipality with a large enough territorial extension to encompass substantial intra-urban temperature variability.

Several extensions to this work are possible. The framework presented can be generalised to different study regions and health outcomes. Extending the analysis to the national scale will require a different spatiotemporal modelling strategy, as the larger spatial domain and high number of monitoring stations may challenge current specifications. A different development will be the propagation of posterior uncertainty from the Bayesian exposure models into the second-stage epidemiological estimation, allowing for coherent joint inference across the two stages of the analysis.

\subsection*{Acknowledgments}
E.C. was supported by the European Union -- NextGenerationEU, under the PNRR project ``C\_PA - DM118 P.A. Pubblica Amministrazione: Metodologie statistiche per il supporto alle decisioni in contesto sanitario pubblico: stima dell'impatto degli eventi climatici estremi sulla salute della popolazione generale e la costruzione di modelli di allerta rapida", CUP: B53C23002660006, Missione 4, Componente 1 (Missione I.4.1, PNRR Scholarships for Public Administration).

E.C. and G.J.L. were partially supported by the Sapienza University project ``Leaving No One Behind: Methods for Sustainability"  ID nr.: RD124190DA1146AA - CUP: B83C24007080005.

J.C.-M. was partially supported by MCIU (PEICTI 2024) under Mobility Grant CAS24/00229 (research visit at Sapienza University of Rome); MCIU/AEI/10.13039/501100011033/FEDER,UE under Grant PID2023-150234NB-I00; and Gobierno de Aragón under Research Group E46\_23R: Modelos Estocásticos.

S.G. was supported by the Doctoral Training Programme ``Science and Solutions for a Changing Planet" Scholarship (SSCP-DTP) provided by the Natural Environment Research Council (NERC) via the Grantham Institute for Climate Change and the Environment, Imperial College London.

M.B. was partially supported by the MRC Centre for Environment and Health, (Grant number: MR/L01341X/1) and the NIHR Health Protection Unit in Chemical and Radiation Threats and Hazards. 

\subsection*{Data and code availability}

Ground monitoring station data are available at \url{https://scia.isprambiente.it/dati-e-indicatori/}. ERA5-Land data can be obtained from the Copernicus Climate Data Store at \url{https://cds.climate.copernicus.eu/datasets/reanalysis-era5-land?tab=overview}. Municipal elevation data are provided by ISTAT at \url{https://www.istat.it/classificazione/principali-statistiche-geografiche-sui-comuni/}. Mortality data are available from Istituto Superiore di Sanità but restrictions apply to the availability of these data, which were used under license for the current study, and so are not publicly available. Data are, however, available from the authors upon reasonable request and with permission of Istituto Superiore di Sanità. Requests for access to the mortality dataset should be directed at: \href{email:giada.minelli@iss.it}{giada.minelli@iss.it}. The code to replicate the analyses is available on GitHub at \url{https://github.com/emilianoceccarelli/temperatures_mortality}.

\subsection*{Author contributions}
E.C., J.C.-M. and M.B conceived the study. E.C. prepared the data and performed the analyses. M.B., G.M. and G.J.L. contributed to the interpretation of the results. E.C. and S.G. drafted the manuscript. M.B., J.C.-M. and G.J.L. revised the manuscript and all authors approved the final version.

\subsection*{Conflict of interest}

The authors declare no potential conflict of interests.

\section*{Supporting information}

The following supporting information is available as part of the online article:

\bibliographystyle{plainnat}  
\bibliography{references}  

\appendix
\section{Additional tables}

\begin{table}[h]
\centering
\begin{tabular}{lc}
  \hline
Spatiotemporal method & Temperature ($^\circ$C) \\ 
  \hline
  Method 1: 0.5 GQRM & 28.07 \\ 
  Method 2: GGPM & 26.94 \\ 
  Method 3: ERA5-Land & 24.82 \\ 
   \hline
\end{tabular}
\caption{MMTs ($^\circ$C) by spatiotemporal methods.}
\label{tab::minRR}
\end{table}


\end{document}